\documentclass[aip,pra,twocolumn,showpacs]{revtex4}
\usepackage{graphicx,color}
\usepackage{amsmath}
\usepackage{longtable}
\begin{document}

\title{Thermodynamics of Yukawa fluids near the one-component-plasma limit}

\author{Sergey A. Khrapak,$^{1,2,}$\footnote{Also at Joint Institute for High Temperatures, Russian Academy of Sciences, Moscow, Russia} Igor L. Semenov,$^1$ L\'{e}na\"{\i}c Cou\"{e}del,$^2$ and Hubertus M. Thomas$^1$ }
\date{\today}
\affiliation{$^1$Forschungsgruppe Komplexe Plasmen, Deutsches Zentrum f\"{u}r Luft- und Raumfahrt,
Oberpfaffenhofen, Germany\\$^2$Aix-Marseille-Universit\'{e}, CNRS, Laboratoire PIIM, UMR 7345, 13397
Marseille cedex 20, France }

\begin{abstract}
Thermodynamics of weakly screened (near the one-component-plasma limit) Yukawa fluids in two and three dimensions is analyzed in detail. It is shown that the thermal component of the excess internal energy of these fluids, when expressed in terms of the properly normalized coupling strength, exhibits the scaling pertinent to the corresponding one-component-plasma limit (the scalings differ considerably between the two- and three-dimensional situations).
This provides us with a simple and accurate practical tool to estimate thermodynamic properties of weakly screened Yukawa fluids. Particular attention is paid to the two-dimensional fluids, for which several important thermodynamic quantities are calculated to illustrate the application of the approach.
\end{abstract}

\maketitle

\section{Introduction}

Thermodynamics of Yukawa systems (particles interacting via the pairwise Yukawa repulsive potential) is of considerable interest, in particular in the context of conventional plasmas, dusty (complex) plasmas, and colloidal suspensions. For small point-like particles the Yukawa potential (also known as screened Coulomb or Debye-H\"{u}ckel potential) reads
\begin{equation}\label{Yukawa}
V(r)= \epsilon (\lambda/r)\exp(-r/\lambda),
\end{equation}
where $\epsilon$ is the energy scale, $\lambda$ is the screening length, and $r$ is the distance between a pair of particles. For charged particles immersed in a neutralizing medium we have $\epsilon\lambda=Q^2$, where $Q$ is the particle charge, and the screening length $\lambda$ is equal to the corresponding Debye radius. Such system are fully characterized by two dimensionless parameters. The first is the coupling parameter, $\Gamma=Q^2/aT$, where $a$ is the characteristic interparticle separation [in this paper we take $a=(4\pi n/3)^{-1/3}$ in three dimensions and $a=(\pi n)^{-1/2}$ in two dimensions, where $n$ is the particle density] and $T$ is the temperature (in energy units). The second is the screening parameter, defined as $\kappa=a/\lambda$.

Statical and dynamical properties of Yukawa systems in both two- and three-dimensional configurations have been extensively investigated using various theoretical and simulation techniques. Thermodynamic properties have also received considerable attention. The accurate results from Monte Carlo (MC) and molecular dynamics (MD) numerical simulations for some thermodynamic functions (in particular, for internal energy and compressibility) have been tabulated for a wide (but discrete) range of state variables $\Gamma$ and $\kappa$~\cite{Meijer,Tejero,Farouki1994,Hamaguchi,CG,Totsuji04}. Recently, simple and accurate expressions to evaluate thermodynamic properties of three-dimensional (3D) Yukawa fluids~\cite{PractUP} and solids~\cite{Solid} have been proposed, which are very convenient for practical applications. These expressions are based on the Rosenfeld-Tarazona (RT) scaling~\cite{Rosenfeld1998} of the excess internal energy in the fluid state (and harmonic approximation in the solid state) and deliver impressively high accuracy in a wide range of the phase diagram of 3D Yukawa systems.

The focus of the present study is on the practical estimation of the thermodynamics of Yukawa fluids near the one-component-plasma (OCP) limit. The OCP corresponds to the limit $\kappa\rightarrow 0$ for Yukawa systems. We show that for $\kappa\lesssim 1$ the OCP scaling of the thermal component of the excess internal energy of 3D Yukawa fluids is superior to that of the RT scaling. This allows us to somewhat improve the accuracy of our previous practical approach to the thermodynamics of 3D Yukawa fluids~\cite{PractUP} in the case of weak screening. More importantly, we show that the thermal component of the excess internal energy exhibits a quasi-universal scaling also in two-dimensional systems of soft repulsive particles (which is, however, quite different compared to the RT or OCP scaling in 3D case). Based on the 2D OCP limit we then produce a simple analytical expression for the excess energy of weakly screened 2D Yukawa systems. The accuracy of this expression is verified by comparison with the results from MD simulations. We demonstrate how this expression can be used to evaluate the specific heat, free energy, compressibility factor, and isothermal compressibility modulus of 2D Yukawa fluids in practical situations.

\section{Yukawa fluids in three dimensions}

The reduced excess energy (excess energy per particle in units of the system temperature) can be conveniently divided into static and thermal components,
\begin{equation}\label{div}
u_{\rm fl}=u_{\rm st}+u_{\rm th}.
\end{equation}
The static contribution corresponds to the value of internal energy when the particles are frozen in some regular structure (e.g. crystalline lattice) and the thermal corrections arise due the deviations from these fixed position due to the particle thermal motion. In the strongly coupled classical OCP, the simplest estimate of the static component can be obtained using the ion sphere model (ISM)~\cite{Baus,Ichimaru}. In this model each particle is restricted to the spherical cell of radius $a$, filled with the neutralizing background. Each cell is quasineutral, the cells do not overlap, and, therefore, $u_{\rm st}$ is simply equal to the electrostatic energy of such cell. The latter can be readily calculated, yielding $u_{\rm st}=-\tfrac{9}{10}\Gamma$~\cite{Ichimaru} (the energy is negative due to particle-background interactions). In fact, this can be mathematically proven to be the exact lower bound on the excess energy of the OCP~\cite{Lieb}. The ISM approach provides a rather good estimate of the actual OCP excess energy at strong coupling, in both fluid and solid phases (for example, the Madelung constant of the OCP forming a body-centered-cubic lattice is $M_{\rm bcc}=-0.895929\Gamma$). Another exact lower bound on the OCP excess energy is provided by the Debye-H\"{u}ckel (DH) construction~\cite{Mermin}. This bound is a reasonable measure of the actual energy at weak coupling. Recently a physically motivated interpolation between the DH and ISM limits has been discussed in Ref.~\cite{Hybrid3D}.

\begin{figure}
\includegraphics[width=7.5cm]{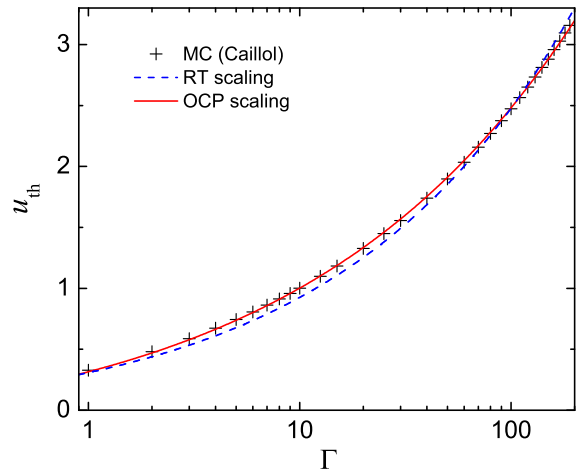}
\caption{(Color online) Thermal component of the reduced excess energy, $u_{\rm th}$, of the strongly coupled 3D OCP fluid versus the coupling parameter $\Gamma$. Crosses correspond to the exact data from MC simulations~\cite{Caillol99}. The (red) solid curve is the OCP fit of Eq.~(\ref{3Dfit}). The (blue) dashed curve corresponds to the RT scaling, given by Eq.~(\ref{RT}).}
\label{fig1}
\end{figure}

When the value of the static component of the excess energy is specified (ISM approach in our case), the thermal component is also well defined. Based on the accurate Monte-Carlo (MC) simulation results from Ref.~\cite{Caillol99}, a simple two-term expression has been proposed~\cite{OCP2014}
\begin{equation}\label{3Dfit}
u_{\rm th}=0.5944\Gamma^{1/3}-0.2786.
\end{equation}
This fit, along with the MC numerical data, is plotted in Fig.~\ref{fig1}. The exponent $s=1/3$ (or close to that) in Eq.~(\ref{3Dfit}) provides particularly good agreement with the numerical data. This fact has been documented in a number of previous studies~\cite{Caillol99,OCP2014,Stringfellow,Farouki1994,Dubin1999}.

For Yukawa systems, the reduced excess energy can also be decomposed into the static and thermal components [Eq.~(\ref{div})], with the static component evaluated using the generalization of the ISM to these systems (see, in particular, Ref.~\cite{ISM} for the details). Note that, in contrast to the OCP limit where the neutralizing background must be present to keep thermodynamic functions finite and well defined, for model Yukawa systems two realizations are possible. The first corresponds to the conventional system of charged particles immersed into neutralizing medium, which provides exponential screening of the particle-particle electrical interactions and keeps the overall charge neutrality (this situation can to some extent mimic real colloidal suspensions and complex plasmas~\cite{IvlevBook}). The second is the imaginary one-component system of particles interacting via the Yukawa potential without any neutralizing medium, which we will refer to as the single component Yukawa system (SCYS), and which is merely used as a useful model in condensed matter research. The difference between the excess energies of these two systems is related to the interactions between the particles and the neutralizing medium. The difference can be evaluated as~\cite{ISM,Ham94}
\begin{equation}\label{Umedium}
u_{\rm m}=-\frac{3\Gamma}{2\kappa^2}-\frac{\kappa\Gamma}{2},
\end{equation}
where the first term represents the energy of the medium that, on average, neutralizes the
charge of the particles (it exactly compensates the energy of the particle-particle interactions when the particles are at complete disorder). The second term gives the energy of the sheath around each particle.
The static component of the internal energy in the ISM approximation is~\cite{ISM}
\begin{equation}\label{u_st}
u_{\rm st}(\kappa,\Gamma) = \frac{\kappa(\kappa+1)\Gamma}{(\kappa+1)+(\kappa-1)e^{2\kappa}}-\frac{\kappa\Gamma}{2}-\frac{3\Gamma}{2\kappa^2}.
\end{equation}
In view of Eq.~(\ref{Umedium}), the first term of the above equation describes the static energy of the SCYS. The coefficient of proportionality, $M_{\rm fl}=u_{\rm st}/\Gamma$, can be  referred to as the fluid Madelung constant~\cite{Rosenfeld2000}. For SCYS we have therefore
\begin{equation}\label{M_fl}
M_{\rm fl}(\kappa) = \frac{\kappa(\kappa+1)}{(\kappa+1)+(\kappa-1)e^{2\kappa}}.
\end{equation}
Note, that the same result [Eq.~(\ref{M_fl})] can be obtained using the Percus-Yevick (PY) radial distribution function of hard spheres in the limit $\eta =1$, where $\eta$ is the hard sphere packing fraction~\cite{Rosenfeld2000}.

To the best of our knowledge, Rosenfeld and Tarazona were first to discuss a quasi-universal behavior of the thermal component of the internal energy ($u_{\rm th}=u_{\rm ex}-M_{\rm fl}\Gamma$) for a wide class of fluids formed by soft repulsive particle (in particular, they focused on inverse-power-law and Yukawa potentials) in three dimensions~\cite{Rosenfeld1998,Rosenfeld2000}. The accuracy of this scaling for various model systems has been recently investigated in extensive numerical simulations~\cite{Ingebrigtsen}. It has been observed that the RT scaling is particulary useful for liquids with strong correlations
between equilibrium fluctuations of virial and potential energy~\cite{Ingebrigtsen}, which are referred to as Rosklide-simple or just Rosklide systems~\cite{BacherNature}. OCP and strongly coupled Yukawa systems belong to this class and, thus, RT scaling can be quite useful to describe thermodynamic properties of these systems.

Previously, a variant of the RT scaling has been successfully used to obtain practical expressions for the internal energy and pressure of 3D Yukawa fluids with $\kappa\lesssim 5$ across coupling regimes~\cite{PractUP}.
The expression for the thermal component of the excess energy suggested in this work is
\begin{equation}\label{RT}
u_{\rm th}\simeq 3.2(\Gamma/\Gamma_{\rm m})^{2/5} - 0.1,
\end{equation}
where $\Gamma_{\rm m}$ is the coupling parameter at the fluid-solid phase transition. More recently, it has been also pointed out that for Yukawa fluids sufficiently close to freezing an even simpler scaling, $u_{\rm th}\simeq 3.1(\Gamma/\Gamma_{\rm m})^{2/5}$, is more appropriate~\cite{Solid}.

The first question to be addressed now is which of the scalings, OCP [Eq.~(\ref{3Dfit})] or RT [Eq.~(\ref{RT})] is more appropriate in the OCP limit. Figure~\ref{fig1} shows the comparison of these scaling with the exact MC simulation  data from Ref.~\cite{Caillol99}. To plot the RT scaling we have used $\Gamma_{\rm m}\simeq 171.8$ for the OCP in three dimensions, as reported in Ref.~\cite{Hamaguchi96} (the true $\Gamma_{\rm m}$ for the OCP can be slightly higher, around $\simeq 174$~\cite{OCP2014,Dubin1999}; we use $\Gamma_{\rm m}\simeq 171.8$ here for the sake of consistency with the Yukawa case, see below). Figure~\ref{fig1} demonstrates that the RT scaling describes fairly well the numerical data, but the OCP scaling is somewhat more accurate, especially in the regime $1\lesssim \Gamma\lesssim 100$. This is absolutely not surprising, since the OCP scaling is nothing but the best simple fit to the numerical data shown in Fig.~\ref{fig1}.

\begin{figure}
\includegraphics[width=7.5cm]{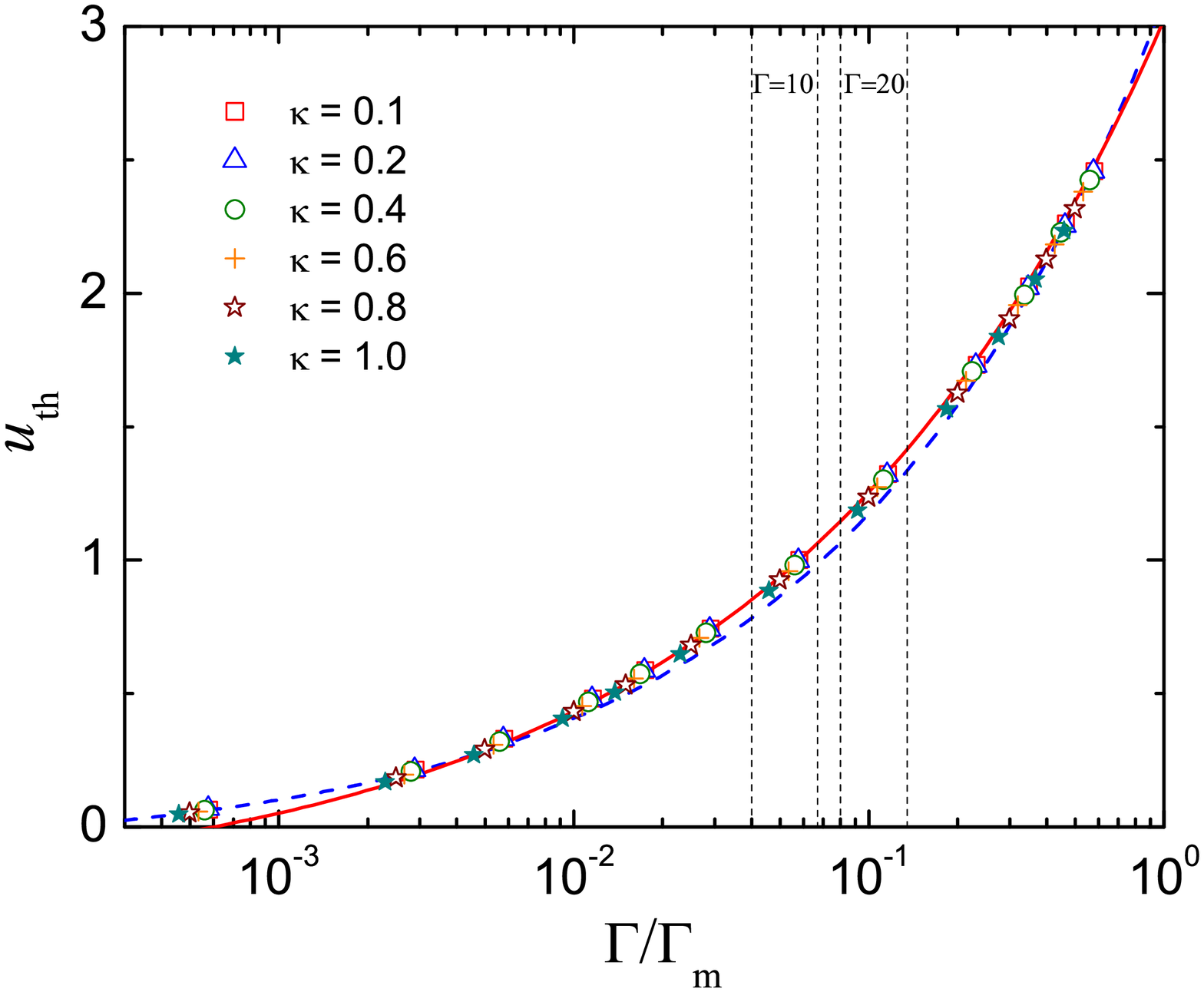}
\caption{(Color online) Thermal component of the reduced excess energy of weakly screened ($\kappa\leq 1$) Yukawa fluids in three dimensions  as a function of the reduced coupling parameter $\Gamma/\Gamma_{\rm m}$. Symbols are the MC simulation results from Ref.~\cite{CG}. The (red) solid curve shows the OCP scaling based on Eq.~(\ref{3Dfit}) and the (blue) dashed curve corresponds to the RT scaling, given by Eq.~(\ref{RT}). The vertical lines denote the two groups of symbols, corresponding to $\Gamma=10$ and $\Gamma=20$, where the deviation between these two scalings is most pronounced.}
\label{fig2}
\end{figure}

The second relevant question is which of the scalings is more relevant for Yukawa systems near the OCP limit (i.e., at weak screening). To answer this, we have used the numerical data for the excess energy of Yukawa fluids tabulated in Ref.~\cite{CG}, subtracted the static contribution [Eq.~(\ref{u_st})], and plotted the remaining thermal component as a function of $\Gamma/\Gamma_{\rm m}$ in Fig.~\ref{fig2}. To do that we used the fit for $\Gamma_{\rm m}(\kappa)$, proposed in Ref.~\cite{Hamaguchi96}:
\begin{equation}
\Gamma_{\rm m}(\kappa) \simeq 171.8 + 42.46\kappa^2 + 3.841\kappa^4,
\end{equation}
which is applicable for $\kappa\lesssim 1.4$. The corresponding numerical data for $\kappa\leq 1.0$ are shown in Fig.~\ref{fig2} along with the OCP and RT scalings. We see that the OCP scaling is a better approximation for $\Gamma/\Gamma_{\rm m}\gtrsim 2\times 10^{-3}$. The deviations between these two scalings are particularly observable for intermediate coupling $\Gamma\sim {\mathcal O}(10)$, as indicated in Fig.~\ref{fig2}. The observation that the OCP scaling provides better description for the thermal component of the excess energy of weakly screened ($\kappa\lesssim 1$) Yukawa systems allows us to improve the previous practical approach to the thermodynamics of Yukawa fluids in this regime. The modifications to the expressions derived in Ref.~\cite{PractUP} are straightforward and we do not discuss this further in detail, except for a relevant comment below.

\begin{figure}
\includegraphics[width=7.5cm]{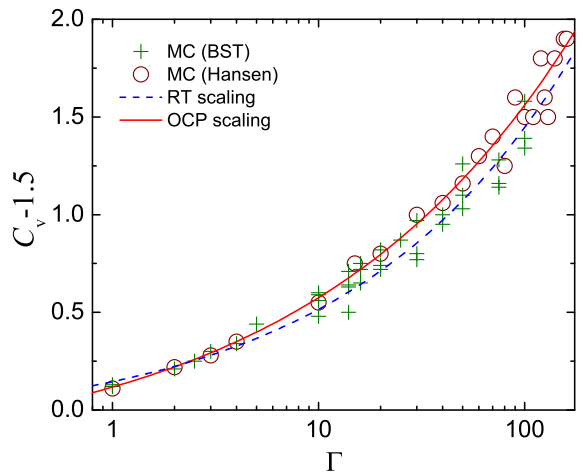}
\caption{(Color online) Reduced excess isochoric heat capacity $c_{\rm v}$ of the one-component-plasma in 3D versus the coupling parameter $\Gamma$. Crosses are the MC results by Brush, Sahlin, and Teller (BST)~\cite{Brush}. Different crosses for the same $\Gamma$ correspond to the systems with different number of particles and configurations (for details see Ref.~\cite{Brush}). Circles denote the MC results by Hansen~\cite{Hansen73}. The (red) solid curve is based on the OCP scaling of Eq.~(\ref{3Dfit}) and the (blue) dashed curve corresponds to the RT scaling of Eq.~(\ref{RT}).   }
\label{fig3}
\end{figure}

It is clear that the thermodynamic functions depending on the total excess energy (like e.g. pressure and compressibility modulus), will be only slightly affected by the differences between the OCP and RT scalings of the thermal energy, since the static energy component dominates for soft repulsive potentials. However, other thermodynamic functions can be more sensitive to these differences. For example, the reduced specific heat at constant volume is determined by the reduced thermal energy alone,
\begin{equation}\label{Cv3D}
c_{\rm v}(\kappa,\Gamma)=\tfrac{3}{2}+u_{\rm th}-\Gamma(\partial u_{\rm th}/\partial \Gamma).
\end{equation}
In Figure~\ref{fig3} we plot the available numerical data for $c_{\rm v}$ for the OCP, along with the analytical expressions based on the OCP and RT scalings. The scattering of the numerical data points does not allow us  to clearly discriminate between the OCP and RT scalings. Nevertheless, based on the comparison shown in Figs.~\ref{fig1} and \ref{fig2}, it is not unreasonable to expect that the OCP scaling is more appropriate for very soft repulsive potentials, which implies, for instance, $c_{\rm v}^{\rm ex}\sim T^{-1/3}$, instead of $c_{\rm v}^{\rm ex}\sim T^{-2/5}$ (according to the RT scaling) in this regime.

\section{Yukawa fluids in two dimensions}

As in the previous section, we first consider the OCP limit. In two-dimensions, two different systems are actually referred to as the OCP. The first is characterized by logarithmic interactions, $V(r)=-Q^2\ln(r/L)$, where $L$ is an arbitrary length scale~\cite{Totsuji79,Caillol82,Leeuw82}. The logarithmic potential emerges from the solution of the two-dimensional Poisson equation around the central point-like particle~\cite{Totsuji79}.
In this case the system is conventionally characterized by the coupling parameter $\Gamma=Q^2/T$, which does not depend on the particle density. In the second system, the interaction is of conventional Coulomb-type, $V(r)=Q^2/r$, but the interacting particles are confined to the 2D plane. In this second case, the coupling parameter is defined similarly to the OCP (or Yukawa) system in 3D, $\Gamma=Q^2/aT$, with  $a$ being the 2D analog of the Wigner-Seitz radius (defined in the Introduction).

\begin{figure}
\includegraphics[width=7.5cm]{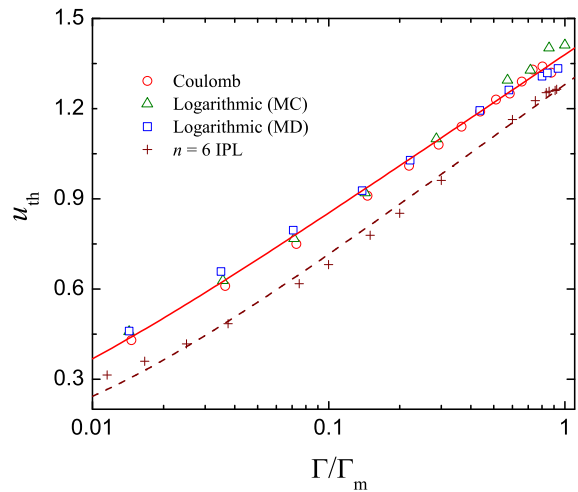}
\caption{(Color online) Thermal component of the reduced excess energy of several soft repulsive systems in two dimensions as a function of the reduced coupling parameter $\Gamma/\Gamma_{\rm m}$. Circles correspond to the numerical data for the Coulombic interaction in 2D~\cite{Gann79}, triangles show the MC results for logarithmically repelling particles confined to the surface of a sphere~\cite{Caillol82},  while squares correspond to the MD
results for logarithmic repulsion on a plane~\cite{Leeuw82}. The crosses correspond to the IPL potential $V(r)\propto r^{-n}$ with $n=6$~\cite{Allen83}. The (red) solid curve shows the OCP scaling for two dimensional systems, Eq.~(\ref{2Dfit}). The brown curve is the fit to the $n=6$ IPL data using the same functional form of Eq.~(\ref{2Dfit}).}
\label{fig4}
\end{figure}

The numerical results for the internal energy of both OCP systems are available in the literature~\cite{Caillol82,Leeuw82,Totsuji78,Gann79}. The first interesting point to address, is wether the scaling of the thermal component of the excess energy in two-dimensions is also quasi-universal for soft repulsive systems (like the two OCPs considered here), similarly to the 3D case. For the static component of the energy in the 2D case we use the triangular lattice sums (Madelung energies), which are $u_{\rm M}=-0.37438\Gamma$ for the logarithmic and $u_{\rm M}=-1.106103\Gamma$ for the Coulomb potential, respectively~\cite{Caillol82,Gann79}. (The ion disc model, an analog of the ISM in 3D, can be constructed for logarithmic interactions~\cite{Caillol82}, but we are not aware of any such construction for the Coulomb interaction in 2D). Then, subtracting the static component from the full excess energy (when necessary) we can obtain the thermal energy component. The resulting dependence of $u_{\rm th}$ on the reduced coupling parameter $\Gamma/\Gamma_{\rm m}$ is shown in Figure~\ref{fig4}. To produce this figure we assumed $\Gamma_{\rm m}= 140$ for the logarithmic interaction~\cite{OCP2014,Caillol82,Leeuw82} and $\Gamma_{\rm m}=137$ for the Coulomb interaction~\cite{Hartmann05}. The closeness of $\Gamma_{\rm m}$ values for these two systems is likely a coincidence, since the physical meaning of the coupling parameters is quite different for the logarithmic and Coulomb interactions.

The numerical data for the thermal component of the excess energy for 2D OCP systems shown in Fig.~\ref{fig4} tend to collapse on a single quasi-universal curve, which is analogous to the OCP scaling in three dimensions. Some scattering of the data points is present, but no clear systematic trend is observed, indicating that this can simply reflect the level of accuracy of the simulation results. The dependence of $u_{\rm th}$ on $\Gamma/\Gamma_{\rm m}$ has a logarithmic character (in contrast to the power-law scaling in 3D) and we found convenient to approximate the data by the following functional dependence:
\begin{equation}\label{2Dfit}
u_{\rm th}\simeq b_1\ln\left[1+b_2(\Gamma/\Gamma_{\rm m})\right],
\end{equation}
with the coefficients $b_1= 0.231$ and $b_2=391.655$. Also shown in Fig.~\ref{fig4} are the numerical data for the thermal energy of the fluid with inverse power law (IPL) interaction, $V(r)=\epsilon(\sigma/r)^{n}$, with $n=6$~\cite{Allen83}. To produce this plot we have used the notation $\Gamma\equiv (\epsilon/T)(\sigma/a)^{n}$, and took $\Gamma_{\rm m}\simeq 105.9$ at the fluid-solid phase transition (in fact, this is the fluid-hexatic transition)~\cite{Kapfer}. We see that the universal scaling only holds for sufficiently soft interactions, which is not the case for the IPL $n=6$ case (the thermal energy is systematically lower). However, the fit of the form (\ref{2Dfit}) can still be reasonably applied, see the dashed curve in Fig.~\ref{fig4} (the fit yields $b_1=0.251$ and $b_2=163.290$ for the $n=6$ IPL).

Now, by analogy with approach discussed in the previous section, we apply the observed soft-repulsion scaling of $u_{\rm th}$ to evaluate the excess energy of weakly screened Yukawa systems in 2D geometry. The static component of the excess energy is associated with the triangular lattice sum (Madelung energy), which for the Yukawa interaction can be approximated by~\cite{Totsuji04}
\begin{equation}\label{Madelung2D}
u_{\rm M}\simeq -1.1061\Gamma\left(1-0.4555\kappa+0.0999\kappa^2-0.0087\kappa^3\right).
\end{equation}
Note that this static energy accounts for particle-particle correlations and the energy of the neutralizing medium (this is why the energy is negative). The energy of the neutralizing medium (plasma) compensates the energy coming from particle-particle interactions when no correlations are present. It can be therefore calculated using the virial energy equation with the radial distribution function $g(r)\equiv 1$, that is
\begin{equation}
u_{\rm pl} = -\frac{\pi n}{T}\int_0^{\infty}rV(r)dr=-\frac{\Gamma}{\kappa},
\end{equation}
where $V(r)$ is the Yukawa potential. The term $-u_{\rm pl}$ is sometimes referred to as the positive Hartree part, while the total energy is then called the correlational part~\cite{Hartmann05}. Obviously, the energy of the SCYS in two dimensions can be obtained as a sum of the Hartree and the correlational parts. (The total energy may also include the contribution from particle-sheath interactions, similarly to the 3D case. However, this term does not affect main thermodynamic quantities of interest and we drop it from further consideration).

The dependence $\Gamma_{\rm m}(\kappa)$ for two dimensional Yukawa systems has been approximated in Ref.~\cite{Hartmann05} by the following fit
\begin{equation}\label{Melting2D}
\Gamma_{\rm m}(\kappa)\simeq \frac{131}{1-0.388\kappa^2+0.138\kappa^3-0.0138\kappa^4}.
\end{equation}
This fit underestimates by $\simeq 4\%$ the transitional coupling parameter of the OCP, but describes relatively well the melting points found from the bond angular correlation analysis (see Fig.~6 of Ref.~\cite{Hartmann05}) up to $\kappa = 3.0$.

\begin{figure}
\includegraphics[width=7.0cm]{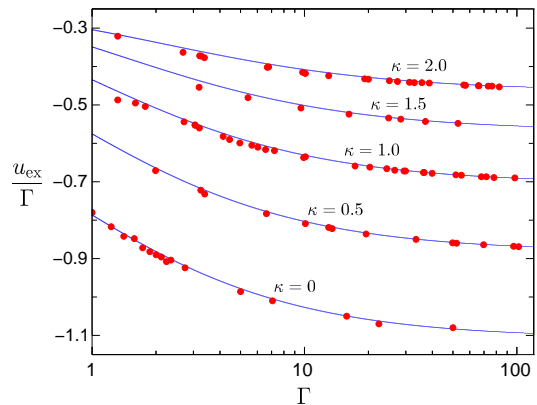}
\caption{(Color online) Reduced excess energy $u_{\rm ex}/\Gamma$ as a function of the coupling parameter $\Gamma$ for weakly screened two dimensional Yukawa fluids. The symbols correspond to the numerical results from Ref.~\cite{Totsuji04}. The solid curves represent our approximation. Shown are the results for five screening parameters:  $\kappa=0.0,~0.5,~1.0,~1.5$, and $2.0$. }
\label{fig5}
\end{figure}

The combination of Eqs.~(\ref{2Dfit}) - (\ref{Melting2D}) allows us to calculate the energy of strongly coupled Yukawa fluids in 2D. We compare our results with the energy data tabulated in Ref.~\cite{Totsuji04}, which are in good agreement with the subsequent simulations in Ref.~\cite{Hartmann05}. The comparison is shown in Figure~\ref{fig5}. The agreement is rather good, the maximum relative deviation from the numerical results does not exceed few percent and is much smaller than that for the majority of the data points.

\begin{table*}
\caption{\label{Tab1} Excess energy of the two-dimensional single component Yukawa fluid in the vicinity of the fluid-solid transition. Three coupling regimes, characterized by $\Gamma/\Gamma_{\rm m}=0.8$, $0.9$, and $0.95$ have been investigated [$\Gamma_{\rm m}$ has been evaluated according to Eq.~(\ref{Melting2D})]. For each of these regimes four values of the screening parameter $\kappa$ have been chosen (first column). The results from numerical MD simulations (MD) are compared with the theoretical prediction (Theory). }
\begin{ruledtabular}
\begin{tabular}{lcccccc}
 & \multicolumn{2}{c}{$\Gamma/\Gamma_{\rm m}=0.80$}  &  \multicolumn{2}{c}{$\Gamma/\Gamma_{\rm m}=0.90$}  &   \multicolumn{2}{c}{$\Gamma/\Gamma_{\rm m}=0.95$}  \\ \cline{2-7}
 $\kappa$ & MD & Theory & MD & Theory & MD & Theory \\ \hline
0.5 & $128.95 \pm 0.34$ & 128.93 & $144.82\pm 0.35$ & 144.90 & $152.79\pm 0.36$ & 152.89  \\
1.0 & $43.62\pm 0.10$ & 43.59 & $48.88\pm 0.12$ & 48.90  &  $51.55 \pm 0.13$  & 51.55\\
1.5 & $21.42\pm 0.05$ & 21.43 & $23.95 \pm 0.06$ & 23.97 & $25.24 \pm 0.06$  & 25.24 \\
2.0 & $12.89\pm 0.03$ & 12.99 & $14.36 \pm 0.03$ & 14.48 & $15.09\pm 0.04$  & 15.22\\
\end{tabular}
\end{ruledtabular}
\end{table*}

The data tabulated in Ref.~\cite{Totsuji04} do not cover the regime of very strongly coupled fluids, near the fluid-solid phase transition. We, therefore, performed additional simulations for 2D Yukawa fluids near freezing.  Simulations have been performed on graphics processing unit (NVIDIA GTX 960) using the hoomd-blue software~\cite{HB1,HB2}. We used $N=160000$ Yukawa particles in a rectangular box with the periodic boundary conditions. The cutoff radius for the potential has been chosen $L_{\rm cut}\simeq 26.6\lambda$. Simulations have been performed in the canonical ensemble ($NVT$) with the Langevin thermostat at a temperature corresponding to the desired target coupling parameter $\Gamma$. We first equilibrated the system for one million time steps and measured and averaged the excess energy for another million of steps. The results of this simulation along with their comparison with the theoretical predictions is summarized in Table~\ref{Tab1}.  The agreement between the theory and numerical experiment is excellent for $\kappa\leq 1.5$. In this regime the deviations between the theoretical and MD results are all within the simulation uncertainty. For $\kappa=2.0$, the theory systematically overestimates the excess energy obtained in simulations, but the relative difference is below $1\%$. This qualitative trend is not unexpected. As the potential softness decreases, the OCP scaling tends to overestimate the actual thermal component of the excess energy. An example, corresponding to $n=6$ IPL potential, is shown in Fig.~\ref{fig4}.

Summarizing this Section, we have proposed simple and reliable practical method to estimate thermodynamics of weakly screened Yukawa fluids in two dimensions. The accuracy of this method is very good, at least for $\kappa\lesssim 2$. In the next sections we provide an exemplary calculation of several thermodynamic functions.

\section{Thermodynamic functions of two-dimensional Yukawa fluids}

We use the conventional thermodynamic identities~\cite{LL} to calculate several useful quantities for Yukawa fluids in two dimensions. In doing so it is convenient to operate with the Yukawa system phase state variables $\kappa$ and $\Gamma$. If the density and temperature of the neutralizing medium are kept fixed, we have $\Gamma\propto (aT)^{-1}\propto T^{-1}n^{1/2}$ and $\kappa\propto a\propto n^{-1/2}$. This implies
\begin{displaymath}
\frac{\partial \Gamma}{\partial T}=-\frac{\Gamma}{T}, \quad \frac{\partial \Gamma}{\partial n}=\frac{1}{2}\frac{\Gamma}{n}, \quad \frac{\partial \kappa}{\partial T}= 0, \quad \frac{\partial \kappa}{\partial n}=-\frac{1}{2}\frac{\kappa}{n}.
\end{displaymath}
The practical model expression for $u_{\rm ex}(\kappa,\Gamma)$ can then be employed to calculate various thermodynamic quantities.

\begin{figure}
\includegraphics[width=8cm]{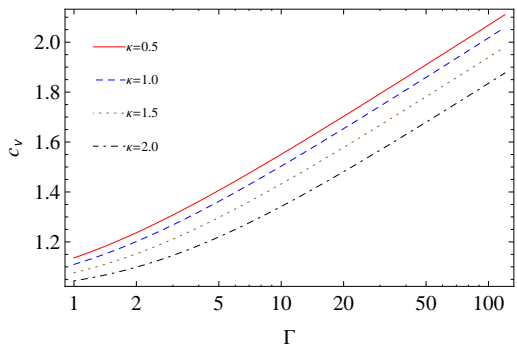}
\caption{(Color online) Reduced isochoric heat capacity $c_{\rm v}$ of the two-dimensional weakly screened Yukawa fluid as a function of the coupling parameter $\Gamma$. The curves denote the calculation using the present    approximation for four values of the screening parameter: $\kappa=0.5,~1.0,~1.5$, and $2.0$.  }
\label{fig6}
\end{figure}

The first example deals with the reduced isochoric heat capacity, which is related to the excess energy via $c_{\rm v}=2+u_{\rm ex}-\Gamma(\partial u_{\rm ex}/\partial \Gamma)$. Since the static contribution is linear in $\Gamma$, $c_{\rm v}$ is completely determined by the thermal component of the excess energy, see Eq.~(\ref{Cv3D}). We have calculated the dependence of $c_{\rm v}$ on $\Gamma$ for several values of $\kappa$ ($\kappa\leq 2.0$) and plotted the resulting curves in Fig.~\ref{fig6}.

\begin{figure}
\includegraphics[width=8cm]{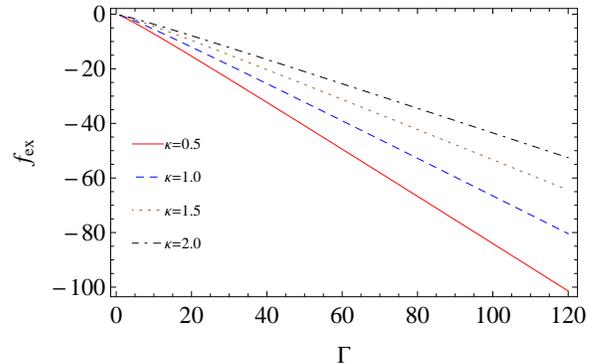}
\caption{(Color online) Reduced excess free energy $f_{\rm ex}$ of the two-dimensional weakly screened Yukawa fluid as a function of the coupling parameter $\Gamma$. The notation is the same as in Fig.~\ref{fig6}.}
\label{fig7}
\end{figure}

The second quantity we consider is the reduced excess free energy of Yukawa fluids. It is related to the excess energy by a standard integration,
\begin{equation}\label{free}
f_{\rm ex}(\kappa,\Gamma)=\int_0^{\Gamma}d\Gamma'u_{\rm ex}(\kappa, \Gamma')/\Gamma'.
\end{equation}
An exact result would be obtained if the dependence $u_{\rm ex}(\kappa,\Gamma)$ is known in the entire range of coupling, starting from $\Gamma=0$. This is not the case here, because we have proposed an approximation for the sufficiently strong coupling regime (say $\Gamma>1$). The conventional procedure can be applied to start the integration from $\Gamma=1$ and add the weak coupling correction $f_{\rm ex}(\kappa, 1)$. The latter can be calculated using the analytical results available in the literature (see e.g.~\cite{Totsuji04}). However, this contribution is vanishingly small at $\Gamma\gg 1$ and can be omitted. Since both $u_{\rm st}$ and $u_{\rm th}$ are linear in $\Gamma$ in the weak coupling limit, we will not introduce serious errors by starting the integration in Eq.~(\ref{free}) from $\Gamma'=0$. The obtained results for the reduced free energy are shown in Fig.~\ref{fig7}. These compare very favorably with the data presented in Fig.~3 of Ref.~\cite{Totsuji04} (including the regime $\Gamma\lesssim 1$ shown in the inset of this figure).

\begin{figure}
\includegraphics[width=7.5cm]{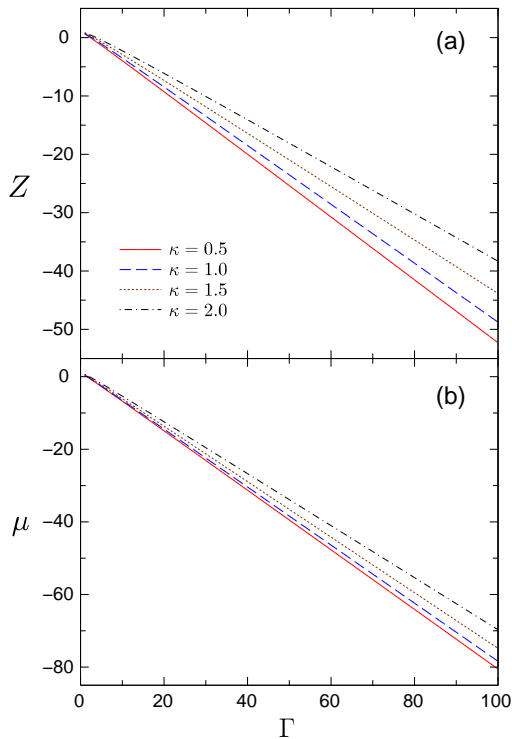}
\caption{(Color online) Compressibility factor $Z$ (a) and the isothermal compressibility modulus $\mu$ (b) of the two-dimensional weakly screened Yukawa fluid as functions of the coupling parameter $\Gamma$. The notation is the same as in Fig.~\ref{fig6}. }
\label{fig8}
\end{figure}

Finally, we calculate the compressibility factor (reduced pressure),
\begin{equation}
Z(\kappa,\Gamma)=1+\frac{\Gamma}{2}\frac{\partial f_{\rm ex} }{\partial \Gamma}-\frac{\kappa}{2}\frac{\partial f_{\rm ex}}{\partial \kappa},
\end{equation}
and the isothermal compressibility modulus,
\begin{equation}
\mu(\kappa,\Gamma)=Z+(\Gamma/2)(\partial Z/\partial \Gamma)-(\kappa/2)(\partial Z/\partial \kappa).
\end{equation}
These quantities can be of considerable interest in the context of wave phenomena in Yukawa fluids~\cite{DAV}. The calculated $Z$ and $\mu$, as functions of $\Gamma$, are plotted in Fig.~\ref{fig8} for several values of the screening parameter $\kappa$. Our present calculation for the compressibility factor compares very favorably with the MD simulation results presented in Fig.~4 of Ref.~\cite{Totsuji04} (again, including the regime $\Gamma\lesssim 1$ shown in the inset of this figure). Thus, simple and accurate practical tool to estimate thermodynamic properties of 2D Yukawa fluids is now available.

\section{Conclusion}

The Rosenfeld-Tarazona (RT) scaling~\cite{Rosenfeld1998,Rosenfeld2000} of the thermal component of the excess energy of simple fluids with soft repulsive interactions is a useful tool to estimate their thermodynamic properties.
Previously, the approach based on the RT scaling arguments was applied to three-dimensional Yukawa fluids and very good accuracy was documented~\cite{PractUP,Solid}. In this paper we have shown that there is still room for some slight improvements in the regime of weak screening. In this regime, the one-component-plasma-like scaling of the thermal energy is superior to the RT scaling and thus even higher accuracy in estimating thermodynamic properties can be reached. Moreover, we have documented the existence of similar scaling (quasi-universal behavior) of the thermal energy for two-dimensional fluids with soft repulsive interactions (the functional form of the scaling varies significantly between the two- and three-dimensional cases). Based on that the total excess energy of weakly screened two-dimensional Yukawa fluids has been evaluated and good agreement with both the existing and new results from direct molecular dynamics simulations has been observed. To further demonstrate the application of this approach, various important thermodynamic quantities of the two-dimensional Yukawa fluids have been evaluated in a wide range of coupling.

\begin{acknowledgments}
At a late stage of this work SAK has been supported by the A*MIDEX grant (Nr.~ANR-11-IDEX-0001-02) funded by the French Government ``Investissements d'Avenir'' program.
\end{acknowledgments}


\begin{thebibliography}{99}

\bibitem{Meijer} E. J. Meijer and D. Frenkel, J. Chem. Phys. {\bf 94}, 2269 (1991).
\bibitem{Tejero} C. F. Tejero, J. F. Lutsko, J. L. Colot, and M. Baus, Phys. Rev. A {\bf 46}, 3373 (1992).
\bibitem{Farouki1994} R. T. Farouki and S. Hamaguchi, J. Chem. Phys. {\bf 101}, 9885 (1994).
\bibitem{Hamaguchi} S. Hamaguchi, R. T. Farouki, and D. H. E. Dubin, Phys. Rev. E {\bf 56}, 4671 (1997).
\bibitem{CG} J. M. Caillol and D. Gilles, J. Stat. Phys. {\bf 100}, 933 (2000).
\bibitem{Totsuji04} H. Totsuji, M. S. Liman, C. Totsuji, and K. Tsuruta, Phys. Rev. E {\bf 70}, 016405 (2004).

\bibitem{PractUP} S. A. Khrapak and H. Thomas, Phys. Rev. E {\bf 91}, 023108 (2015).
\bibitem{Solid} S. A. Khrapak, N. P. Kryuchkov, S. O. Yurchenko, and H. M. Thomas, J. Chem. Phys. (2015, in press).

\bibitem{Rosenfeld1998} Y. Rosenfeld and P. Tarazona, Mol. Phys. {\bf 95}, 141 (1998).

\bibitem{Baus} M. Baus and J. P. Hansen, Phys. Rep. {\bf 59}, 1 (1980).
\bibitem{Ichimaru} S. Ichimaru, Rev. Mod. Phys. {\bf 54}, 1017 (1982).

\bibitem{Lieb} E. H. Lieb and H. Narnhofer, J. Stat. Phys. {\bf 12}, 291 (1975).
\bibitem{Mermin} N. D. Mermin, Phys. Rev. {\bf 171}, 272 (1968).
\bibitem{Hybrid3D} S. A. Khrapak and A. G. Khrapak, Phys. Plasmas {\bf 22}, 044504 (2015).

\bibitem{Caillol99} J. M. Caillol, J. Chem. Phys. {\bf 111}, 6538 (1999).

\bibitem{OCP2014} S. A. Khrapak and A. G. Khrapak, Phys. Plasmas {\bf 21}, 104505 (2014).

\bibitem{Stringfellow} G. S. Stringfellow, H. E. DeWitt, and W. L. Slattery, Phys. Rev. A {\bf 41}, 1105 (1990).

\bibitem{Dubin1999} D. H. E. Dubin and T. M. O'Neil, Rev. Mod. Phys. {\bf 71}, 87 (1999).

\bibitem{ISM} S. A. Khrapak, A. G. Khrapak, A. V. Ivlev, and H. M. Thomas, Phys. Plasmas {\bf 21}, 123705 (2014).

\bibitem{IvlevBook} A. Ivlev, H. L\"{o}wen, G. Morfill, and C. P. Royall, {\it Complex Plasmas and Colloidal Dispersions: Particle-resolved Studies of Classical Liquids and Solids} (World Scientific, Singapore, 2012).

\bibitem{Ham94} S. Hamaguchi and R. T. Farouki, J. Chem. Phys. {\bf 101}, 9876 (1994).

\bibitem{Rosenfeld2000} Y. Rosenfeld, Phys. Rev. E {\bf 62}, 7524 (2000).

\bibitem{Ingebrigtsen} T. S. Ingebrigtsen, A. A. Veldhorst, T. B. Schr{\o}der, and J. C. Dyre, J. Chem. Phys. {\bf 139}, 171101 (2013).
\bibitem{BacherNature} A. K. Batcher, T. B. Schr{\o}der, and J. C. Dyre, Nature Communications {\bf 5}, 5424 (2014).

\bibitem{Hamaguchi96} S. Hamaguchi, R. T. Farouki, and D. H. E. Dubin, J. Chem. Phys. {\bf 105}, 7641 (1996).

\bibitem{Brush} S. G. Brush, H. L. Sahlin, and E. Teller, J. Chem. Phys. {\bf 45}, 2102 (1966).
\bibitem{Hansen73} J. P. Hansen, Phys. Rev. A {\bf 8}, 3096 (1973).

\bibitem{Totsuji79} H. Totsuji, Phys. Rev. A {\bf 19}, 2433 (1979).
\bibitem{Caillol82} J. M. Caillol, D. Levesque, J. J. Weis and J. P. Hansen, J. Stat. Phys. {\bf28}, 325 (1982).
\bibitem{Leeuw82} S. W. de Leeuw and J. W. Perram, Physica A {\bf113}, 546 (1982).

\bibitem{Totsuji78} H. Totsuji, Phys. Rev. A {\bf17}, 399 (1978).
\bibitem{Gann79} R. C. Gann, S. Chakravarty and G. V. Chester, Phys. Rev. B {\bf20}, 326 (1979).

\bibitem{Hartmann05} P. Hartmann, G. J. Kalman, Z. Donko, and K. Kutasi, Phys. Rev. E {\bf 72}, 026409 (2005).

\bibitem{Allen83} M. P. Allen, D. Frenkel, W. Gignac, and J. P. McTague, J. Chem. Phys. {\bf 79}, 4206 (1983).

\bibitem{Kapfer} S. C. Kapfer and W. Krauth, Phys. Rev. Lett. {\bf 114}, 035702 (2015).

\bibitem{HB1} HOOMD-blue web page: http://codeblue.umich.edu/hoomd-blue
\bibitem{HB2} J. A. Anderson, C. D. Lorenz, and A. Travesset,  J. Comput. Phys. {\bf 227}, 5342 (2008).

\bibitem{LL} L. D. Landau and E. M. Lifshitz, {\it Statistical Physics} (Elsevier, 2005).

\bibitem{DAV} S. A. Khrapak and H. M. Thomas, Phys. Rev. E {\bf 91}, 033110 (2015).

%
%



%

%
%

%
%
%
%
%
%
%
%
%

\end{thebibliography}
\end{document}